\documentstyle[prl,aps,multicol,subfigure,psfig,subeqnar]{revtex}

\begin{document}

\title{Bose-Einstein condensates in standing waves:
The cubic nonlinear Schr\"odinger equation with a periodic potential}
\author{Jared C. Bronski$^{1}$, Lincoln D. Carr$^{2}$\cite{byline},
Bernard Deconinck$^{3}$, and J. Nathan Kutz$^{3}$\\}
\address{$^{1}$Department of Mathematics, University of Illinois 
         Urbana-Champaign, Urbana, IL 61801, USA\\}
\address{$^{2}$Department of Physics, University of Washington, 
         Seattle, WA 98195-1560, USA\\}
\address{$^{3}$Department of Applied Mathematics, University of 
         Washington, Seattle, WA 98195-2420, USA\\}
\maketitle

\date{\today}

\begin{abstract}
We present a new family of stationary solutions to the cubic nonlinear
Schr\"odinger equation with a Jacobian elliptic function potential.  In the limit
of a sinusoidal potential our solutions model a dilute gas Bose-Einstein
condensate trapped in a standing light wave.  Provided the ratio of the
height of the variations of the condensate to its DC offset is small
enough, both trivial phase and nontrivial phase solutions are shown to be
stable.  Numerical simulations suggest such stationary states are
experimentally observable.
\end{abstract}


\begin{multicols}{2}

The dilute--gas Bose-Einstein condesate (BEC) in the quasi--one--dimensional
regime is modeled by the cubic nonlinear Schr\"odinger equation (NLS) with
a potential~\cite{dalfovo1,carr22,key1}.  The various
traps which are used to contain the BEC have spurred the solution of the
NLS with new potentials~\cite{kunze1,carr15}. BECs trapped in a standing
light wave have been used to study phase coherence~\cite{anderson3} and
matter-wave diffraction~\cite{ovchinnikov1} and have been predicted to have
applications in quantum logic~\cite{jaksch1} and matter-wave
transport~\cite{choi1}.  Exact solutions have been obtained for the
Kronig-Penney potential~\cite{barra1} and some researchers have used a Bloch
function description~\cite{steel1}.  In this letter, we study new
explicit solutions of the NLS with a Jacobian elliptic function potential.

We consider the mean-field model of a 
quasi-one-dimensional repulsive BEC trapped in an
external potential which is given
by the nonlinear Schr\"odinger equation~\cite{dalfovo1} 
\begin{equation}
\label{eqn:NLS}
 i\psi_t = -\frac{1}{2}\psi_{xx} + |\psi|^2 \psi 
        + V(x) \psi \, .
\end{equation}
In experiments, the trapping potential is generated by a standing 
light wave~\cite{anderson3}.  As a model for such a potential
we use the periodic potential
\begin{equation}
    V(x) = -V_0~ {\rm sn}^2(x,k)
\end{equation}
where ${\rm sn}(x,k)$ denotes the Jacobian elliptic 
sine function~\cite{abro} with elliptic modulus $0\leq k\leq 1$.  In the limit 
$k\rightarrow 1^-$, $V(x)$ becomes an array of well-separated 
hyperbolic secant potential barriers or wells, while in the limit 
$k\rightarrow 0^+$ it becomes purely sinusoidal.  We note that
for intermediate values (e.g. $k=1/2$) the potential closely resembles
the sinusoidal behavior and thus provides a good approximation
to the standing wave potential generated experimentally~\cite{anderson3}.

We present stationary solutions in closed form and study their stability 
analytically and numerically.
We begin by constructing solutions to
Eq.~(\ref{eqn:NLS}) which have the form
\begin{equation}
\label{eqn:solns1}
\psi(x,t) = r(x)\exp(i (\Theta(x)-\omega t))
\end{equation}
where
\begin{subeqnarray}
\label{eqn:solns2}
 r^2(x)   &=& (V_0+k^2)~ {\rm sn}^2(x,k)+ B \\ 
 \Theta(x) &=& c \displaystyle{\int_0^x \frac{dx'}{r^2(x')}} \\
 \omega   &=& \displaystyle{\frac{1}{2} \left( 1+ k^2 + 3B -
                 \frac{BV_0}{k^2+V_0} \right) } \\
 c^2      &=& \displaystyle{B \left(1+\frac{B}{k^2+V_0}\right)
                \left(k^2+V_0+Bk^2\right)}
\end{subeqnarray}
where $B$ determines a mean amplitude and acts as a DC
offset for the number of condensed atoms.  The strength of the nonlinearity,
which for the BEC is a function of both the atomic coupling and the number of
condensed atoms, is determined by the parameters $V_0+k^2$ and $B$, as is
apparent in the amplitude of the solutions given by Eq.~(\ref{eqn:solns1}).
Note that if $x$ is scaled so that $V(x)$ undergoes only a single oscillation
on the ring (in the limit $k\rightarrow 1$) the Jacobian elliptic potential provides
a model of a single barrier or well~\cite{carr24}.  For
simplicity we focus on two special cases: (1) $k$ arbitrary and trivial phase
($c=0$), and (2) $k=0$ with nontrivial phase ($c\neq 0$) so that the solutions
are trigonometric functions.
%

{\em Trivial Phase Case --} In the limit of $c=0$, the solutions 
given in Eqs.~(\ref{eqn:solns1})-(\ref{eqn:solns2}) reduce to 
\begin{equation}
  \label{eqn:elliptic1}
  \psi(x,t)=\sqrt{V_0+k^2}~{\rm sn}(x,k)\exp[-i(1+k^2)t/2]
\end{equation}
valid for $V_0\ge -k^2$, and
\begin{subeqnarray}
  \label{eqn:elliptic2}
&& \hspace*{-.35in} \psi(x,t) \!=\! \sqrt{-(V_0 \!+\! k^2)}~{\rm cn}(x,k)
   \exp[-i(V_0\!+\!k^2\!-\!\frac{1}{2})t], \\
&& \hspace*{-.35in} \psi(x,t) \!=\! \sqrt{-(1\!+\!\frac{V_0}{k^2})}~{\rm dn}(x,k)
   \exp[i(1\!+\!\frac{V_0}{k^2}\!-\!\frac{k^2}{2})t]
\end{subeqnarray}
valid for $V_0\le-k^2$ where cn$(x,k)$ and dn$(x,k)$ are Jacobian elliptic
functions~\cite{abro}.  These solution branches 
%
%
\begin{figure}[t]
\centerline{\psfig{figure=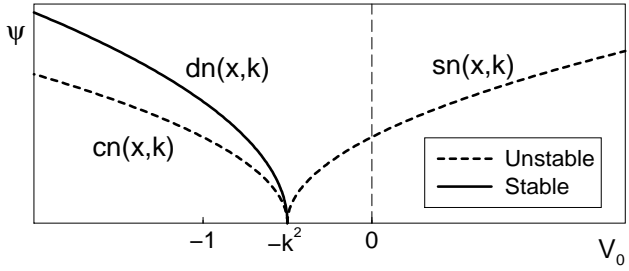,width=83mm}}
\begin{center}
\begin{minipage}{83mm}
\caption{Regions of validity for trivial phase Jacobian elliptic solutions.  The
sn$(x,k)$ and cn$(x,k)$ branches are found to be unstable while
the dn$(x,k)$ branch is stable.\label{fig:bif}}
\end{minipage}
\end{center}
\end{figure}
%
%
\noindent
are illustrated
in Fig.~1 along with their stability properties found
from analytic calculations and observed in numerical simulation.

We can prove that the dn$(x,k)$ branch of solutions is linearly
stable.  To do so, we linearize about the dn$(x,k)$ solution
branch given by Eq.~(\ref{eqn:elliptic2}b) so that
\begin{equation}
   \psi(x,t)= \left( \phi_0(x) + \phi(x,t) \right) \exp(-i\omega t)
\end{equation}
where $\phi_0(x)\exp(-i\omega t)$ 
is the exact solution given by Eq.~(\ref{eqn:elliptic2}b)
and $\phi(x,t)\ll 1$ is a small perturbation to the exact solution.
This leads to the following linearized eigenvalue problems
\begin{subeqnarray}
\label{eqn:linearized}
   L_- L_+ R &=& \lambda^2 R \\
   L_+ L_- I &=& \lambda^2 I
\end{subeqnarray}
where $\phi(x,t)=(R+iI)\exp(i\lambda t)$ is
decomposed into its real and imaginary parts.  The operators
$L_-$ and $L_+$ are both self-adjoint and periodic differential
operators:
\begin{subeqnarray}
   L_- &=& \frac{1}{2} \partial_x^2 - \phi_0(x)^2 - V(x) - \omega \\
   L_+ &=& \frac{1}{2} \partial_x^2 - 3\phi_0(x)^2 - V(x) - \omega \, . 
\end{subeqnarray}
Thus the eigenvalue problem closely resembles that of soliton
solutions of the NLS~\cite{wein} with the additional difficulty
of $V(x)$ being a periodic potential.
   
We are able to prove stability for the whole dn$(x,k)$ branch of
solutions and instability for the cn$(x,k)$ and sn$(x,k)$ branches
near their emanation point $V_0=-k^2$ (see Fig.~1).  We note that
stability or instability is proven by showing that the eigenvalue
$\lambda^2$ in Eq.~(\ref{eqn:linearized}) is positive or negative
respectively.  For the dn$(x,k)$ branch of solutions, we find
that $L_+$ is a positive definite operator.  This allows us 
to construct its inverse.  We then consider the related
problem $L_- I=\lambda^2 L_+^{-1} I$ which is self-adjoint
with respect to the weighting operator $L_+^{-1}$.  This then
allows us to show that $\lambda^2>0$ so that the eigenvalues 
remain real, establishing stability.

%
\begin{figure}
\centerline{\psfig{figure=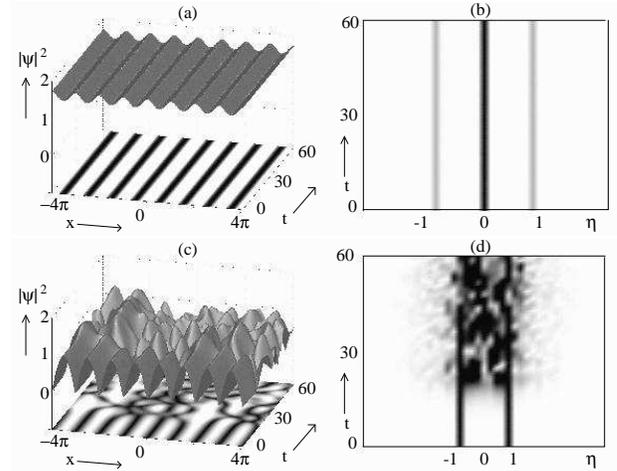,width=83mm}}
\begin{center}
\begin{minipage}{83mm}
\caption{
Evolution of initially perturbed Jacobian elliptic dn$(x,k)$ and sn$(x,k)$
solutions.  Panels (a)--(b) correspond to the 
stable dn$(x,k)$ branch of solutions of Fig.~1 with $k=1/2$ and
$V_0=-1$.  The wavenumber spectrum ($\eta$) remains constant and 
consists of the zero mode and wavenumber one.  These reflect the
DC offset and strength of oscillation, respectively.  Panels 
(c)--(d) demonstrate the instability of the sn$(x,k)$ solution with
$k=0$ and $V_0=1$.  Here the instability is seen to develop
around wavenumber unity as predicted analytically.  The
cn$(x,k)$ branch of solutions exhibits the same instability.
\label{fig:elliptic}
}
\end{minipage}
\end{center}
\end{figure}
%

In contrast to the dn$(x,k)$ solutions, numerical experiments 
suggest that all cn$(x,k)$ and sn$(x,k)$ solutions are linearly unstable.
Although we have been unable 
to show this for the whole solution branch, we 
have shown this perturbatively for $V_0+k^2<\!\!<1$ using the same
technique as outlined above for the dn$(x,k)$ branch of solutions. 
In this limit, the dispersion 
relation for small disturbances near $V_0+k^2<\!\!<1$ is
\begin{equation}
  \omega^2 = \frac{(\eta^2-1)^2}{2} + C(V_0+k^2)\frac{\eta^2-1}{2},
\end{equation}
where $\eta$ is the wavenumber of the disturbance
and $C$ is a non-zero constant. From this it follows that
there is a band of unstable modes which occurs  
near $\eta =1$. Numerical experiments show that 
this unstable band persists for all $V_0$ and grows 
as $V_0+k^2$ becomes large. We emphasize that the 
stable dn$(x,k)$ branch of solutions consists of oscillations
about a non-zero mean value, while the unstable cn$(x,k)$ and sn$(x,k)$ branches
have zero mean, suggesting that the offset has an important effect on  
the stability properties.  Rigorous proofs of the linear stability calculations 
will be considered elsewhere~\cite{carr24}.

Figure~\ref{fig:elliptic} illustrates numerical solutions of
Eq.~(\ref{eqn:NLS}) for initial \hfill conditions \hfill
consisting \hfill of the exact sn$(x,k)$ and  
%
%
\begin{figure}
\centerline{\psfig{figure=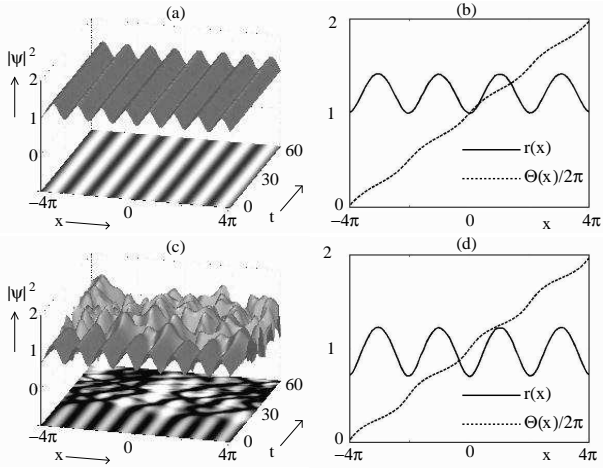,width=83mm}}
\begin{center}
\begin{minipage}{83mm}
\caption{Evolution of perturbed trigonometric
solutions ($k=0$) with nontrivial phase.  In panels (a)--(b), $B=1$ so that
the solution is dn$(x,k)$--like, leading to a stable evolution.  
The initial amplitude
and phase are depicted in the top right.  For $B=1/2$,
the solution is unstable, as seen in (c).  Its initial phase and amplitude
are depicted in panel (d).  This indicates that
a sufficiently high DC offset is required to stabilize 
the evolution.
\label{fig:trig}
}
\end{minipage}
\end{center}
\end{figure}
%
%
\noindent
dn$(x,k)$
solutions perturbed with a small amount of initial stochastic white noise,
along with the corresponding Fourier spectra. 
Figures~\ref{fig:elliptic}(c) and \ref{fig:elliptic}(d) depict
the evolution of a sn$(x,k)$ initial condition perturbed by noise.  For
this simulation, $V_0=1$ and $k=0$ so that the solution is
a simple sinusoid (see Fig.~1).  The solution is
unstable, and diverges rapidly from the exact solution given in
Eq.~(\ref{eqn:elliptic2}). From the plot of the Fourier spectrum, 
it is clear that the instability begins in a neighborhood of wavenumber 
$\eta=1$ as predicted from the linear stability analysis.  Similar
behavior is observed for the cn$(x,k)$ branch of solutions. 
Figures~\ref{fig:elliptic}(a) and \ref{fig:elliptic}(b) 
depict the evolution of the stable dn$(x,k)$ branch of
solutions which is initially perturbed by 
white noise. In this simulation $V_0=-1$ and $k=1/2$.
This solution branch is stable, and stays close to the exact
solution for all time.  Note that the Fourier spectrum is dominated by
the zero mode, which determines the DC offset, and the wavenumber one,
which determines the oscillation strength.
 
{\em The Nontrivial Phase, Trigonometric Limit -- }
For $k=0$, ${\rm sn}(x,0)=\sin(x)$. 
The governing evolution Eq.~(\ref{eqn:NLS}) reduces to
\begin{equation}
i\psi_t = -\frac{1}{2}\psi_{xx} + |\psi|^2 \psi - V_0 \sin^2(x) \psi \, .
\label{eqn:nlsekzero}
\end{equation}
Note that by the trigonometric identity 
$2\sin^2(x)=1-\cos(2x)$ this potential is sinusoidal.  
The solutions reduce to 
\begin{equation}
\label{eqn:ctrig}
 \psi(x,t) \!=\! \sqrt{V_0\sin^2 x +B} \exp[i (\Theta(x)\!-\!(1/2+B)t)] 
\end{equation}
where
\begin{equation}
     \tan(\Theta(x))=\pm\sqrt{1+V_0/B}\tan(x)
\end{equation}
determines the nontrivial phase provided that $B\geq-V_0$ for $V_0<0$
and $B\geq 0$ for $V_0>0$.

The results of the previous section imply that the trivial phase solutions
without offset are unstable.  Oscillations about some mean value
is qualitatively similar to a dn$(x,k)$ solution, which suggests that
solutions with sufficiently large $B$ might be stable.  Oscillations
with $|B|$ small are qualitatively like cn$(x,k)$ and sn$(x,k)$
and are expected to be unstable.  Numerical experiments confirm
this.  We note that a linear stability analysis in this
case is more complicated than for the trivial phase case since
the linearized operators do not decouple as in Eq.~(\ref{eqn:linearized}).

In Fig.~\ref{fig:trig} we depict the evolution of a pair of initial conditions
of the form given by Eq.~(\ref{eqn:ctrig}) plus initial white noise.  
Figures~\ref{fig:trig}(c) and \ref{fig:trig}(d) depict the initial 
amplitude and phase along with the evolution of the density $|\psi(x,t)|^2$
for the parameter values $V_0=1.0, B=0.5$. It is clear from the graph that
this stationary solution is unstable.  Figures~\ref{fig:trig}(a) and
\ref{fig:trig}(b) depict the initial
amplitude and 
phase and the evolution of the density for parameter values $V_0=1.0,
B=1.0$. In this case the solution appears stable.  The solution
with the perturbed initial condition stays close to the stationary solution
for long times, which for our time scaling is longer than the lifetime of
typical trapped BECs~\cite{carr22}.

The numerical and analytical results imply that 
in order to obtain a stable condensate it is necessary to have 
solutions which are sufficiently in the nonlinear regime.  
To quantify this, we note 
that from Eqs.~(\ref{eqn:nlsekzero}) and~(\ref{eqn:ctrig}), the number
of particles per well $n$ is given by
$n=(\int_0^\pi |\psi(x,t)|^2 dx)/\pi=V_0/2 + B$.  
In the context of the BEC, and for a fixed atomic coupling strength,
this means a sufficient number of condensed atoms  
per well $n$ is required to provide a DC offset on the order of 
the potential strength.  This ensures stabilization of the condensate.

\begin{figure}
\centerline{\psfig{figure=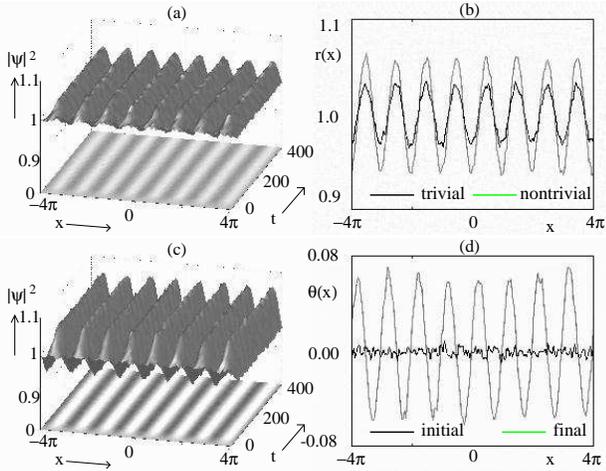,width=83mm}}
\begin{center}
\begin{minipage}{83mm}
\caption{
Stable growth of periodic condensate with DC offset of one 
with (a) initial trivial phase and (c) nontrivial 
phase.  A $\sin^2(x)$ potential was 
adiabatically grown from $V_0=0$ at $t=0$ to $V_0=0.5$ at $t=50$.  
Figure~(b) shows the difference in the final
state of the system given an initial trivial phase (black line) 
and initial nontrivial phase (gray line).  The development of phase structure
on an initial linear phase profile is depicted in (d). 
Here the initial linear phase profile has been subtracted out to give
the initial (black line) and final (gray line) 
phase profiles.
\label{fig:trig2}
}
\end{minipage}
\end{center}
\end{figure}

To demonstrate the physical viability of our results, we
perform numerical simulations consistent with recent
experiments on the BEC~\cite{anderson3}.  
In particular,
we consider a fixed number of condensed atoms with an
initially constant DC profile.
We adiabatically ramp up the periodic 
potential linearly from zero to a fixed potential strength $V_0$.  
In an experiment, the condensate initially has a trivial
phase profile and the solution deforms as depicted in
Fig.~\ref{fig:trig2}(a).  This adiabatic process 
generates a stable solution which is close to the dn$(x,k)$ 
solution branch given by Eq.~(\ref{eqn:elliptic2}b) and depicted
in Fig.~\ref{fig:elliptic}(a).  In contrast, an initial DC profile with
a nontrivial \hfill phase \hfill profile is only stable provided the DC
%
%
%
%
\noindent
offset is sufficiently large in comparison to
the potential strength $V_0$.
For $V_0=0.5$ and an initial DC offset of one, the adiabatically
grown solution is stable with 
larger amplitude fluctuations than those of the 
trivial phase case (Fig.~\ref{fig:trig2}b).  As the condensate
evolves, the initial linear phase profile is deformed
as shown in 
Fig.~\ref{fig:trig2}(d), where the initial linear phase profile
is subtracted out from the initial and final phases. 
This phase deformation is necessary in order for the solution to
remain stationary.  In particular, a linear phase profile
induces a group-velocity shift in the direction of growing phase.
This is in contrast to phase jumps or transition regions which can
cause motion opposite the direction of a positive jump~\cite{carr22,carr15}.  
The stationary solutions generated here suggest that these two opposite
actions effectively balance each other in order for the solutions
to remain localized and stationary in their respective troughs. 
Finally, we note that for $V_0=1$ the potential is sufficiently 
strong so as to destabilize the adiabatically grown solution.  
The instability mechanism
is similar to that observed in Fig.~\ref{fig:trig}(c). 

{\em Conclusions -- } We have considered the repulsive NLS equation with a
Jacobian elliptic function potential as a model for a trapped, quasi-one-dimensional
Bose-Einstein condensate, and thereby produced a whole family of solutions.
We have shown the stability of the dn$(x,k)$ branch of solutions rigorously. 
In addition, we have used a perturbative argument to show the 
instability of the sn$(x,k)$ and cn$(x,k)$ solutions.  In the sinusoidal 
limit we have provided numerical evidence that
the non-trivial phase solutions are stable for sufficiently large
offset $B$.  Stable trivial and nontrivial phase trigonometric 
solutions can be obtained through adiabatic ramping of the potential 
strength provided there is a sufficient DC offset in the number of 
condensed atoms.  As this models a Bose-Einstein
condensate trapped in a standing light wave, our results imply that
sufficiently large number of condensed atoms (DC offset) 
are required to form a stable, periodic condensate.

We benefited greatly from discussions with Ricardo Carretero-Gonzal\'ez, Keith
Promislow, and William Reinhardt.  The work of J. Bronski, L. D. Carr, B.
Deconinck, and J. N. Kutz was supported by National Science Foundation Grants
DMS-9972869, CHE97-32919, DMS-0071568, and DMS-9802920 respectively.


\begin{thebibliography}{10}
\bibitem[*]{byline} to whom correspondence should be addressed
\end{thebibliography}

\end{multicols}
\end{document}